\documentclass[10pt]{article}
\usepackage{amssymb}
\usepackage{amsthm}
\usepackage{authblk}
\usepackage{setspace}
\usepackage{hyperref}
\usepackage[textwidth=6.5in,textheight=8.25in,centering]{geometry}
\usepackage[pdftex]{graphicx}

\onehalfspace
\begin{document}
\begin{center}
{\LARGE{\textbf{Three-fold Constructive Perturbation for Significant Enhancement in Field Emission from Nickel Oxide Nano-Thorn}}}

\vspace{0.5 cm}
\textit{Suryakant Mishra$^a$, Priyanka Yogi$^{a}$, Shailendra K. Saxena$^{a}$,  J. Jayabalan$^{b, c}$, Pankaj R. Sagdeo$^a$ and Rajesh Kumar$^a$}\footnote{Corresponding author email: rajeshkumar@iiti.ac.in} \footnote{\href{http://magse.webs.com/} {\textbf{http://magse.webs.com/}}}

\vspace{0.5 cm}
$^a$ Material Research Laboratory, Discipline of Physics \& MEMS, Indian Institute of Technology Indore, Simrol-453552, Madhya Pradesh, India

$^b$ Nano Materials Laboratory, Materials Science Section, Raja Ramanna Centre for Advanced Technology, Indore - 452013, India

$^c$ Homi Bhabha National Institute, Training School Complex, Anushakti Nagar, Mumbai - 400094, India.


\vspace{1 cm}
ABSTRACT

\end{center}
A power efficient and stable field emission (FE) has been reported here from Nickel Oxide nanostructures. Modification in device geometry and surface micro- (nano-) structure has been found helpful in addressing the bottlenecks in achieving an efficient FE . In terms of threshold and turn on fields, three orders of magnitude better electron FE has been observed in the nickel oxide nanopetals (NiO-NPs) fabricated using simple hydrothermal technique. Uniform and vertically aligned NiO-NPs structures, grown on very flat conducting surface (FTO coated glass), show sharp needles like structures on the top edges of the flakes. These ultrafine structures play the main role in field emission to start at such a low turn on fields. The FE data (J-E plot) has been fitted with Fowler-Nordheim (FN) equation to estimate threshold field value and field enhancement factor which are found to be 3 V/mm and $\sim 5 \times 10^6$ respectively. 
\vspace{0.5cm}

\textbf{Keywords:}  Field emission, Nanotechnology, NiO

\section{Introduction}

 Beginning of the current century has witnessed foundation of great breakthroughs in the field of optics, electronics and optoelectronics achieved by designing novel materials in their nanostructured form [1–-3]. Various nanostructures (NSs) of carbon, silicon and oxides have been studied extensively for application in field emission (FE) in an endeavor to get a FE based display device [4–-8]. Other than the display devices, the FE, a quantum mechanical tunneling phenomena is of great interest, due to its application in devices such as microwave device, x-ray source etc [9,10]. One dimensional (1D) materials having high aspect ratio which provide local electric field enhancement [11,12] providing favorable condition for FE. Nanosheets, a 2D materials, which can get arranged like flower petals (nanopetals [13]),if get well aligned, very good FE is theoretically possible irrespective of the material however, the FE efficiency and ease with which FE can take place may vary and makes a topic of research.
 
Carbon NSs has been established as a landmark material for  FE [4,5,14–-17]. In-spite of being one of the well-studied systems, carbon NSs based FE devices are not available commercially [4]. This has forced scientists to start quest for another material that are good field emitters, parallely, if not as an alternative to the carbon based field emitters. Nickel oxide (NiO) has drawn attentions in recent years due to its dielectric properties  attributed to its wide band gap [18]. Vertically aligned cone-shaped NiO nanowires, fabricated using a simple technique, exhibits a field emission properties [19] with threshold field of $\sim$6500 V/mm and turn on field of $\sim$11500 V/mm giving rise to a field enhancement factor of $\sim$ 2000.  For wider acceptability in FE application a better turn-on \& threshold fields are required and methods to achieve the same should be investigated. Recent study reveals the role of surface morphology in the FE properties from CdS NSs [20]along with nanostructures’ size \& aspect ratio [4,12].

The current letter reports the observation of significant enhancement in the FE properties from rose petal like NiO NSs covered with ultrathin nanothorn fabricated by hydrothermal method [21]. A three orders of magnitude better turn on field and threshold fields as compared to the previous reports on NiO [19] has been reported here. The modifications induced by morphology (nanothorn covered well aligned NiO NSs) and the design (presence of conducting FTO film) introduces certain perturbations in the system that addresses the bottleneck issues associated with limited FE properties from metal oxide NSs.   
 
\section{Experimental Details}

The NiO NSs sample has been fabricated using hydrothermal process using nickel-nitrite and potassium-persulfate as precursors. A 5 hours of heating at 150$^o$C followed by rinsing in water \& dried. The nanostructures were grown on very flat conducting surface (FTO coated glass) after annealing the above sample for 2 hours at 250$^o$C.  Samples were characterized using XRD (Rigaku SmartLab, Cu-K$_\alpha$ radiation $\lambda$ = 1.54 \AA), SEM (Supra 55 Zeiss), AFM (Bruker), UV-Vis (Agilent  make Cary 60). The field emission characteristics of the NiO-NPs film were measured in a parallel-plate-electrode configuration inside the vacuum chamber at 10$^{-6}$ mbar pressure. During FE measurement, the inter-electrode distance maintained at 400 $\mu$m
\section{Results and Discussion}

The scanning electron microscope (SEM) images in Figure 1a shows very dense rose-petal like structures of NiO. Thickness of these petals is approximately 25-30 nm showing very fine thorn like structures on the top of it. The density and uniformity of the film can be appreciated from the right inset in Figure 1a which shows that hundreds of microns wide area can be uniformly deposited with such kind of nanoflakes covered with nano thorn on the top edge. Cross-sectional view of the NiO NSs can be seen in left inset (Figure 1a). To check the alignment of the petals and uniformity of the film grown on the FTO substrate, we have scratched-off the film from the substrate showing very well grown film which are well connected to the substrate. A surface-plots (Figure 1b), carried out using \textit{ImageJ} software, of a portion of the SEM image shows finer structures on the top edge of the nanopetals.  AFM Images of the samples have also been shown in Figure 1c and 1d. Energy dispersive x-ray (EDX) spectrum (Figure S4) has been used to confirm that the nanostructures are of NiO which is in consonance with the FCC phase of NiO as revealed using XRD  and shows a band gap of 3.6 eV as estimated using  UV-Vis spectroscopy. From the above discussion it has been established that the sample contains FCC-structured, crystalline, dense, well aligned nano-petals of NiO which is covered with very fine (nano-) thorns of size of a few nanometers making it an ideal candidate for electron field emission applications. 

\begin{figure}
\begin{center}
\includegraphics[width=12cm]{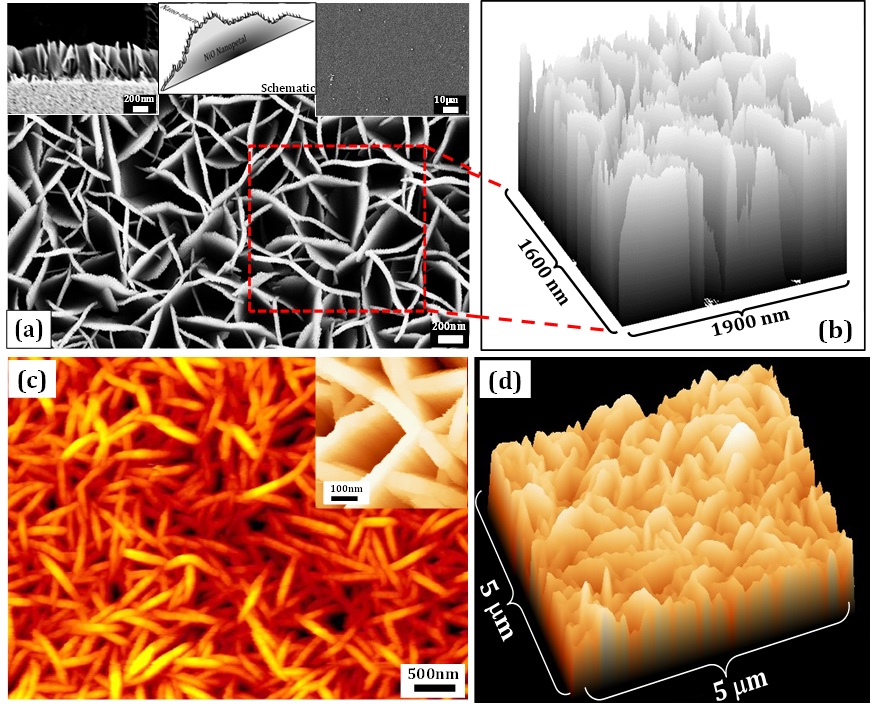}
\caption{SEM micrographs showing (a) Rose-petal type NiO NSs with cross-sectional SEM image (left inset), nanopetals schematic in middle-inset and uniformity of NiO film over hundred microns (right inset).  (b) Surface plot obtained from SEM image using ImageJ. (c, d) 2D and 3D AFM images of the same. }
\end{center}
\end{figure}

The electron FE has been studied from the NiO NSs sample (as cathode) placed in vacuum by placing the counter electrode with suitably applied bias to initiate the FE and has been shown in Figure 2a. The FE from NiO NSs (Figure 2a) have been characterized by measuring the macroscopic current density (J) versus macroscopic electric field. As the bias at NiO-NPs@FTO cold cathode and Cu plate anode increases, it is expected that NPs starts emitting electrons after a threshold value (to be defined later) as a result of tunneling of electrons from the NiO NSs to the cathode through the vacuum gap (0.4 mm in the present case). Figure 2a clearly shows a fairly high noticeable current (mA) flows with an applied bias of millivolts which will be calculated quantitatively later on.  It is apparent from figure 2a that at the electric field of around 3 V/mm sufficiently large numbers of electrons start tunneling out from the surface of cold cathode. 

This electric field appears to be very small as compared to various other nanostructured material  reported so far [22–-24]. Another term important in characterizing the FE properties is the turn on field which is traditionally defined as the field required to get a FE current of 10 $\mu$A/cm$^2$. The observed turn on field is found to be 1.2 mV/$\mu$m (1.2V/mm) which is very less as compared to other NiO NSs [12,19]. The observed turn-on field is at least thousand times (three orders of magnitude) better than the cone shaped NiO nanorods fabricated by Zhang et al [19] using hydrothermal method. Explanation for the observed low turn-on field will be discussed in the later sections. From application point of view, stability is one of the key parameters to be discussed. We have done FE study for multiple cycles as demonstrated in Figure 2a which shows FE in 1$^{st}$, 10$^{th}$, 20$^{th}$, 30$^{th}$ and 40$^{th}$ cycle exhibiting consistency in emission current. In each cycle, emission current retains its value, which is an evidence of cathodic stability.  Furthermore, FE plots in first cycle and 50$^{th}$ cycle have been shown in the insets of Figure 2a separately for comparison. During the cyclic measurement an interesting phenomenon was observed. During the first cyclic scan when voltage goes from initial to final and again traversing back to the initial value hysteresis appears in J-E curve (left inset Figure 2a). It is evident from Figure 2a insets that hysteresis is present in the first cycles but as the cycle proceeds, it disappears without changing the value of turn-on field and corresponding current values. Observed hysteresis are similar to the one observed frequently in carbon based NSs  [25–-28]. 

The FE properties of NiO NSs have also been analyzed using the Fowler-Nordheim (FN) framework by fitting the FE data with the following FN equation (Eq. 1): 
\begin{equation}
J = \frac{A}{\Phi} \beta^2 E^2 e^{-\frac{B \Phi^{\frac{3}{2}}}{\beta E}}
\end{equation}
where A and B are the temperature independent constants with $A=1.54\times 10^{-6} AeV/V^2 and B=6.83 \times 10^3 eV^{-3/2} V/\mu m)$.  $\beta$ is the field-enhancement factor and J, E are the emission current density and applied electric field respectively. Figure 2b shows the FN plot, which is obtained by plotting \textit{ln}(J/E$^2$) vs 1/E for various cycles from the same sample recorded consequently. The little variation in the plots with respect to the cycling number has already been discussed above. It is evident from Eq. 1 that the FN plot is a linear function, when \textit{ln}(J/E$^2$) is plotted as a function of 1/E, with slope equal to  $-\frac{B\Phi^\frac{3}{2}}{\beta}$  whose fitting with the experimental  FE emission data (discrete points in Figure 2b inset) has been used for the calculation of field enhancement factor ($\beta$) and threshold field (E$_{th}$). Linear fitting (solid line in Figure 2b inset) of the FN curve gives a straight line with the slope of 10.66 giving a $\beta$ value of 5713 thousand. It is also clear that intercept of FN equation is a measure of the threshold electric field (E$_{th}$) which comes out to be 2.94 V/mm. The observed enhancement factor and threshold electric field are thousand times better than the reported values [19] for NiO nano rods who has reported $\beta$ $\sim$ 2000 and E$_{th}$ $\sim$ 4000 V/mm. The improved FE properties in the present case are due to the effective improvement in the system due to typical nature of surface morphology containing ultrathin nanothorns present on NiO NSs which is consistent with the ones reported in literature [20] with an improvement in enhancement factor by three orders of magnitude. 

Though it is very enthusiastic to achieve such a low turn-on field, reason for the same must be understood. As the FE properties are dependent on the nanowire diameter and morphology, the likely reason for the low turn-on field may be related to the same in our case also which is supported by the SEM images (Figure 1).  As shown in SEM images (Figures 1), thickness of each petal is around 25--30 nm with sharp needle like edges on the top of it provide higher local electric field enhancement. The extremely fine NSs on the top of NiO nanopetals have multiple advantages that improve the FE properties as explained below. As mentioned earlier, the FE works on the principle of quantum tunneling from the surface through a vacuum gap under the influence of electric field thus depends on the potential barrier and its width. Which means that, tunneling barrier width is one of the important parameters to be addressed if one needs to get FE as per one’s choice. In FE, electrons, having higher energy than the work function, can escape by overcrossing the barrier and those with lower energy needs stronger electric field to overcome the attractive force of the nuclei and tunnel through the barrier  by surpassing the same [29].
 
\begin{figure}
\begin{center}
\includegraphics[width=14cm]{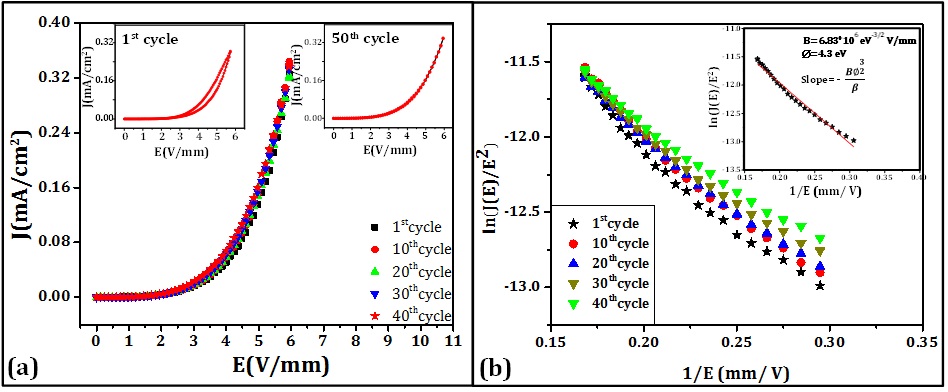}
\caption{(a) Field emission (J-E plot) obtained from NiO-NPs@FTO at various cycles. Insets show J-E curves on 1$^{st}$ and 50$^{th}$ cycle.(b) Fowler-Nordheim fitting of the experimental FE data obtained from the NiO NFs with inset showing the fitting used for $\beta$ calculation.}
\end{center}
\end{figure}

The process of FE and various players involved in the process can be understood using a schematic given in Figure 3 which shows various energy levels and their relative positions/alignment. With no bias applied an additional amount of energy ($\chi$) is required so that electron is free to move. For FE, depending on the applied bias, the effective vacuum level gets modified as shown by dotted lines (for two different field values) for the ideal situations. As a result, the applied bias perturbs the situation in such a way that now a relatively thinner width needs to be tunneled for electron emission. It is also evident from Figure 3 that the width of the potential barrier decreases as a result of increasing field. The actual potential profile, as a result of applied field in such systems, is different from the ideal ones and must be taken into account while understanding the FE properties. The energy state path followed by an electron, emitted by the barrier tunneling as a function of separation (\textit{x}) from cathode [30] is given by:
\begin{equation}
e\Phi = e\Phi_{work} - \frac{e^2}{16\pi\epsilon_0 \chi}-eEx
\end{equation} 
where $e\Phi$ is electric potential energy of an electron $e\Phi_{work}$ and $\epsilon_0$ represent the work function and permittivity of the material respectively. E is the applied electric field during the emission.  Schematic diagrams in Figure 3 shows the energy levels in the absence (E$_0$) and presence of two different electric fields E$_1$ and E$_2$ where E$_1$ $<$ E$_2$  at the interface of vacuum and NiO NSs. Solid green lines (curved) show the typical plot of Eq. 2 for two different values of E chosen arbitrarily. The slanted dotted lines accompanying these green lines represent their corresponding ideal counterpart as discussed above. It is interesting to see that the real situation (green solid lines), obtained using Eq. 2, is more favorable for FE as the barrier height and width both are smaller. It is worth mentioning here that the two situations (corresponding to the two electric fields) can be obtained by using the same bias (voltage) under two different sample morphologies as shown in the diagram next to the green lines. This can be elaborated as follows. The ultrathin nanothorns present in our samples effectively enhance electric field around the finer structures. As a consequence, a given applied bias voltage results in a higher effective electric field for samples containing nanothorns as compared to samples without them.

It is clearly evident from the schematic (Figure 3) that for stronger electric field electrons not only need to be pushed upto a shorter distance to make it free (for emission) but also will require relatively lesser energy to cross the barrier hence resulting in low turn on values. Moreover, electrons enclosed in a strongly confined system (as in nanothorns) will possess more energy as compared to a weekly confined system (as in nanorods of a few tens of nanometers or nanoflakes without nanothorns). Thus the electrons, that already have higher energies, would need further lesser energy to cross the potential barrier before it is attracted by an applied electric field to initiate the FE. This way the morphological variation in our system induced three fold perturbations is responsible for improving the FE from the nanothorn covered NiO nanopetals deposited on conducting FTO film which provides plenty of electrons and further assists in achieving an improved FE from ultrathin NiO NSs.

\begin{figure}
\begin{center}
\includegraphics[width=14cm]{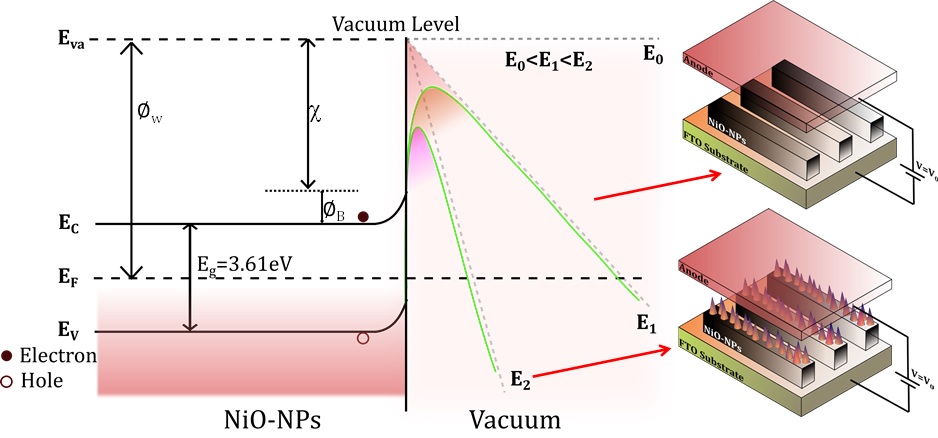}
\caption{Band diagram of NiO-NPs along with favorable bias conditions for tunneling. EC- bottom of conduction band; EF- Fermi level; EV- top of valance band; Actual energy pattern for two different electric fields are represented by solid green lines. Voltage applied for both the conditions is same but resulting in two different effective electric fields due to sample morphological variations as shown in right diagram.}
\end{center}
\end{figure}

\section{Conclusions}

In conclusion, ultrafine nanothorn covered nanopetals of NiO show very efficient field emission properties with three orders of magnitude lower turn on fields. A very stable electron emission at very low threshold field also has been observed from these nanopetals fabricated using simple hydrothermal process. The fabricated NiO nanopetals show well aligned vertical orientation of the sharp edges of these nanostructures with conducting bottom foundation (FTO film) available for biasing as a reservoir of electrons which is helpful in improving the field emission properties. The nanothorns on the nanopetals play a role similar to ``lightening conductors” and thus making field emission favorable by effectively enhancing the electric field around the finer tips with very low applied voltage. As a consequence, decreased tunneling barrier and tunneling width allow electrons, already having relatively higher energy due to quantum confinement effects, to emit under a minimum applied bias. On the whole, modification in device structure and surface micro- (nano-) morphologies allows one to achieve field emission properties at thousand times better electric fields

\subsection*{Acknowledgement} Authors acknowledge financial support from Department of Science and Technology, Govt. of India. Authors are thankful to SIC facility provided by IIT Indore and Mr. Kinny Pandey for his assistance. Author SM is also thankful to MHRD, Govt. of India for providing fellowships. Authors acknowledge Dr. Rama Chari (RRCAT, Indore) for useful discussion and providing AFM facility.

\newpage


\begin{thebibliography}{31}
\bibitem{1}	 Z.W. Pan, Z.R. Dai, and Z.L. Wang, Science 291, 1947 (2001).
\bibitem{2}	 M.H. Huang, S. Mao, H. Feick, H. Yan, Y. Wu, H. Kind, E. Weber, R. Russo, and P. Yang, Science 292, 1897 (2001).
\bibitem{3}	 H. Zeng, W. Cai, P. Liu, X. Xu, H. Zhou, C. Klingshirn, and H. Kalt, ACS Nano 2, 1661 (2008).
\bibitem{4}	 A. Pandey, A. Prasad, J.P. Moscatello, and Y.K. Yap, ACS Nano 4, 6760 (2010).
\bibitem{5}	 A. Pandey, A. Prasad, J.P. Moscatello, M. Engelhard, C. Wang, and Y.K. Yap, ACS Nano 7, 117 (2013).
\bibitem{6}	 H. Gan, H. Liu, Y. Li, Q. Zhao, Y. Li, S. Wang, T. Jiu, N. Wang, X. He, D. Yu, and D. Zhu, J. Am. Chem. Soc. 127, 12452 (2005).
\bibitem{7}	 H. Liu, Q. Zhao, Y. Li, Y. Liu, F. Lu, J. Zhuang, S. Wang, L. Jiang, D. Zhu, D. Yu, and L. Chi, J. Am. Chem. Soc. 127, 1120 (2005).
\bibitem{8}	 H.-C. Wu, T.-Y. Tsai, F.-H. Chu, N.-H. Tai, H.-N. Lin, H.-T. Chiu, and C.-Y. Lee, J. Phys. Chem. C 114, 130 (2010).
\bibitem{9}	 Y. Saito and S. Uemura, Carbon 38, 169 (2000).
\bibitem{10}	 J.-H. Deng, L.-N. Deng, R.-N. Liu, A.-L. Han, D.-J. Li, and G.-A. Cheng, Carbon 102, 1 (2016).
\bibitem{11}	 S. Mishra, P. Yogi, S.K. Saxena, V. Kumar, and R. Kumar, J. Phys. Chem. Lett. 7, 5291 (2016).
\bibitem{12}	 V. Kumar, S.K. Saxena, V. Kaushik, K. Saxena, A.K. Shukla, and R. Kumar, RSC Adv. 4, 57799 (2014).
\bibitem{13}	 S.K. Srivastava, A.K. Shukla, V.D. Vankar, and V. Kumar, Thin Solid Films 492, 124 (2005).
\bibitem{14}	 J.F. Rodriguez-Nieva, M.S. Dresselhaus, and J.C.W. Song, Nano Lett. 16, 6036 (2016).
\bibitem{15}	 J.F. Rodriguez-Nieva, M.S. Dresselhaus, and L.S. Levitov, Nano Lett. 15, 1451 (2015).
\bibitem{16}	 P.S. Weiss, ACS Nano 3, 2434 (2009).
\bibitem{17}	 M.S. Dresselhaus, ACS Nano 4, 4344 (2010).
\bibitem{18}	 G.T. Tyuliev and K.L. Kostov, Phys. Rev. B 60, 2900 (1999).
\bibitem{19}	 Z. Zhang, Y. Zhao, and M. Zhu, Appl. Phys. Lett. 88, 033101 (2006).
\bibitem{20}	 T. Zhai, X. Fang, Y. Bando, Q. Liao, X. Xu, H. Zeng, Y. Ma, J. Yao, and D. Golberg, ACS Nano 3, 949 (2009).
\bibitem{21}	 X. Xia, D. Chao, X. Qi, Q. Xiong, Y. Zhang, J. Tu, H. Zhang, and H.J. Fan, Nano Lett. 13, 4562 (2013).
\bibitem{22}	 G. Rosenman, D. Shur, Y.E. Krasik, and A. Dunaevsky, J. Appl. Phys. 88, 6109 (2000).
\bibitem{23}	 K.K. Naik, R. Khare, D. Chakravarty, M.A. More, R. Thapa, D.J. Late, and C.S. Rout, Appl. Phys. Lett. 105, 233101 (2014).
\bibitem{24}	 Y.W. Zhu, T. Yu, F.C. Cheong, X.J. Xu, C.T. Lim, V.B.C. Tan, J.T.L. Thong, and C.H. Sow, Nanotechnology 16, 88 (2005).
\bibitem{25}	 J. Chen, J. Li, J. Yang, X. Yan, B.-K. Tay, and Q. Xue, Appl. Phys. Lett. 99, 173104 (2011).
\bibitem{26}	 J. Chen, B. Yang, X. Liu, J. Yang, and X. Yan, Appl. Phys. Lett. 108, 193112 (2016).
\bibitem{27}	 E. Cazalas, I. Childres, A. Majcher, T.-F. Chung, Y.P. Chen, and I. Jovanovic, Appl. Phys. Lett. 103, 053123 (2013).
\bibitem{28}	 J. Chen, L. Cui, D. Sun, B. Yang, J. Yang, and X. Yan, Appl. Phys. Lett. 105, 213111 (2014).
\bibitem{29}	 R.H. Fowler and L. Nordheim, Proc. R. Soc. Lond. Ser. Contain. Pap. Math. Phys. Character 119, 173 (1928).
\bibitem{30}	 D.H. Dowell and J.F. Schmerge, Phys. Rev. Spec. Top. - Accel. Beams 12, 074201 (2009).


\end{thebibliography}
\end{document}